\documentclass[aps,prd,twocolumn,showpacs,amsmath,preprintnumbers,nofootinbib,superscriptaddress,floatfix]{revtex4}

\def\scn#1#2{\section{#1}\lb{#2}}
\def\sscn#1#2{\subsection{#1}\lb{#2}}

\usepackage{epsfig,amssymb,amsfonts,verbatim}
\usepackage{bm}
\usepackage{color}

\usepackage{array}

\usepackage[caption=false]{subfig}
\def\sc{0.99} 
\def\sct{0.5} 

\def\bfl{\begin{flushleft}}
\def\efl{\end{flushleft}}
\def\bfr{\begin{flushright}}
\def\efr{\end{flushright}}
\def\bc{\begin{center}}
\def\ec{\end{center}}
\def\be{\begin{equation}}
\def\ee{\end{equation}}
\def\ba{\begin{eqnarray}}
\def\ea{\end{eqnarray}}
\def\baa#1{\begin{array}{#1}}
\def\eaa{\end{array}}
\def\bw{\begin{widetext}}
\def\ew{\end{widetext}}
\def\nn{\nonumber }
\def\lb#1{\label{#1}}
\def\bit{\begin{itemize}}
\def\eit{\end{itemize}}

\def\schrod{Schr\"odinger  }

\def\restm{\tilde{{\rm M}}}
\def\tamden{{\cal M}}

\def\paul{\bm{\sigma}}

\newtheorem{Def}{Def.}[section]

\newtheorem{Prp}[Def]{Proposition}

\def\d{\delta}

\def\e{\epsilon}

\begin{document}

\preprint{\small Phys. Rev. D 94 (2016) 096012 [arXiv:1611.02105]}

\title{
Singularity-free model of electrically charged fermionic particles
and gauged $Q$-balls 
}

\author{Vladimir Dzhunushaliev}
\email{v.dzhunushaliev@gmail.com}
\affiliation{
Department of Theoretical and Nuclear Physics \& IETP, Kazakh National University, 010008 Almaty, Kazakhstan}
\affiliation{
Institute of Systems Science,
Durban University of Technology, 
4000 Durban, South Africa}

\author{Arislan Makhmudov}
\email{arslan.biz@gmail.com}
\affiliation{
Department of Theoretical and Nuclear Physics \& IETP, Kazakh National University, 010008 Almaty, Kazakhstan}

\author{Konstantin G. Zloshchastiev}%
\email{http://bit.do/kgz}
\affiliation{
Institute of Systems Science,
Durban University of Technology, 
4000 Durban, South Africa}

\begin{abstract}
We propose a model of an electrically charged fermion as a regular localized solution of electromagnetic and spinor fields interacting with a physical vacuum, which is phenomenologically described as a logarithmic superfluid. 
We analytically study the asymptotic behavior of the solution, while obtaining its form by numerical methods.
The solution has physically plausible properties, such as finite size, self-energy, total charge and mass. 
In the case of spherical symmetry, its electric field obeys the Coulomb asymptotics at large distances from its core. It is shown that the observable rest mass of the fermion arises as a result of interaction of the fields with the physical vacuum. The spinor and scalar field components of the solution decay exponentially outside the core; therefore they can be regarded as internal degrees of freedom which can only be probed at sufficiently large scales of energy and momentum. 
Apart from conventional Fermi particles, our model can find applications in a theory of exotic localized objects, such as $U(1)$ gauged $Q$-balls with half-integer spin.
\end{abstract}

\pacs{11.27.+d, 11.15.Kc, 11.15.Ex, 03.75.Nt
}

\date{Received: 19 Aug 2016 [PRD], 7 Nov 2016 [arXiv]}
\maketitle

\scn{Introduction}{sec-i}

The models of elementary particles as regular localized solutions arising from suitably chosen field-theoretical models continue to attract substantial interest; a bibliographic and historical overview can be found in Ref. \cite{Dzhunushaliev:2012zb}.
The general consensus is that they can be helpful in understanding two of the oldest, but still relevant, problems in fundamental particle physics: the finite nature of the self-energy of elementary particles, and their possible extendedness \cite{Rajaraman:1982is}. 
These problems are interlinked and manifest themselves in different aspects of quantum field theory and high-energy physics.

In an earlier work \cite{Dzhunushaliev:2012zb}, a relativistic model of an electrical charge as a self-consistent field configuration of an electromagnetic (EM) field interacting with a physical vacuum, effectively described as a logarithmic quantum Bose liquid in the phonon regime, was proposed  \cite{Zloshchastiev:2009aw,Zloshchastiev:2009zwa}.
This liquid was described by the logarithmic \schrod equation which was proposed on other grounds in Refs. \cite{ros69,BialynickiBirula:1976zp}, but for the purposes of quantum gravity and high-energy physics
this equation was used quite recently, see Ref. \cite{Zloshchastiev:2009zw}.
In Ref. \cite{zlo12}, the logarithmic fluid
has been proven to be a proper superfluid, \textit{i.e.}, a quantum Bose liquid that simultaneously possesses the following two properties: 
its spectrum of excitations has the Landau ``roton'' form, which guarantees that dissipation is suppressed at microscopic level, and 
its macroscopic (averaged) equation of state coincides with that of the perfect fluid, at least in the leading approximation. While the liquid itself is non-relativistic, in the linearized (``phononic'') limit, its small excitations, both localized and collective, are known to have the Lorentz symmetry \cite{Zloshchastiev:2009aw,az11}.
This suggested an interesting possibility, that of describing both relativistic particles and gravitational interaction as excitations of a fundamental superfluid background; and it was demonstrated that the dilatonic gravity arises as a low-energy limit of superfluid vacuum theory with logarithmic fluid as a major component of background \cite{Zloshchastiev:2009aw}.
This result was independently confirmed by means of 
the Arnowitt-Deser-Misner approach and generalized Ohta-Mann formalism in Ref. \cite{szm16},
where it was shown that some information about logarithmic superfluid background can be extracted
even in the ``phononic'' (relativistic) limit, by means of geometrical methods
and notions.
Furthermore, there exist some experimental data suggesting a connection between the superfluidity and gravity phenomena \cite{Tajmar:2008aj,Tajmar:2009ar}. 

It was also demonstrated, using different methods, that logarithmic quantum fluids are stable against small perturbations \cite{az11,bo15}.
This stability is an important ingredient for justifying an actual physical existence of such fluids.

Furthermore, in Ref. \cite{Dzhunushaliev:2012zb} it was shown that, in contrast to the EM field propagating in a trivial vacuum, a regular localized solution does exist in the presence of a non-trivial background (which becomes  \textit{de facto} the physical vacuum) represented by the logarithmic superfluid. It was also demonstrated that both the mass and spatial extent of the solution emerge naturally from its self-interaction, also that its charge and energy density distribution acquire a Gaussian-like form. Additionally, it was shown that the solution in the logarithmic superfluid model is stable, unlike the one in the theory of quartic scalar potential.

The goal of this paper is to include half-integer spin into this picture: we are presenting a field-theoretical description of an extended fermionic particle which has both gauge charge and half-integer spin. More specifically, we are proposing a model which consists of a scalar field with logarithmic potential (which represents small fluctuations of the logarithmic superfluid background) interacting with electromagnetic and 
Dirac spinor fields.

Apart from conventional particles, such regular solutions of field-theoretical models can find application in describing exotic localized objects, such as $U(1)$ gauged $Q$-balls \cite{gul14,gul15}.

Surprisingly, our bibliographic research has not found many reports about regular solutions involving spinor fields. 
Initially the nonlinear spinor field, as suggested by the symmetric coupling between nucleons, muons, and leptons, was investigated in a classical approximation in Ref. \cite{Finkelstein1951}.
In Ref. \cite{Rybakov1974}, a nonlinear spinor field equation admitting regular stationary solutions was considered. In that paper, following a causal interpretation of quantum mechanics given by de Broglie in his double solution theory, it was shown that regular solutions can be regarded as describing the internal particle structure. 
Further, the exact static solutions of spinor-field equations with nonlinear terms in the G\"odel universe were obtained in Ref. \cite{shikin1996}, whereas the plane-symmetric solutions of nonlinear spinor field coupled to self-gravitational field were obtained in Ref. \cite{shikin1995}. 
In Ref. \cite{Chakrabarti1988} the exact solutions of the massive Dirac equation were obtained in the $SU(2)$ gauge field background in the Einstein static universe, and in the works \cite{Saha:1996by,Saha:2001ig,Saha:2007if} some other cosmological solutions with spinor fields were found.

The structure of the paper is as follows. The description of the model can be found in Section \ref{sec-mo}, the regular solution and its properties are analyzed in Section \ref{sec-pls}, and  conclusions are drawn in Section \ref{sec-con}.

\scn{The model}{sec-mo}

From now on we adopt the natural units $c=\hbar = \epsilon_0 =1$ and metric signature $(+---)$.

\sscn{Lagrangian and field equations}{sec-mo-b}

We assume the electrodynamic model interacting with a scalar field $\psi$ and spinor field $\chi$,
described by the following Lagrangian
\be
  {\cal L} = -\frac{1}{4} F_{\mu \nu} F^{\mu \nu} +
  \chi^\dagger \gamma^\mu D_\mu \chi +
  \frac{1}{2} \left| D_\mu \Psi \right|^2 -
  V( |\Psi |^2)
	,
\label{1-10}
\ee
where we denoted the covariant derivatives
$D_\mu \Psi = \left( \partial_\mu - i g A_\mu \right) \Psi$
and
$D_\mu \chi = \left(
    i \partial_\mu - e A_\mu
  \right) \chi$,
EM strength tensor
$F_{\mu\nu} = \partial_\mu A_\nu - \partial_\nu A_\mu$, four-dimensional EM potential $A^\mu = \left( \phi, \vec A \right)$ where
$\phi$ and $ \vec A$ are the scalar and vector components, correspondingly.
Besides, here $\dagger$ denotes the Hermitian conjugation, $e$ and $g$ are coupling constants between EM field and spinor and scalar fields, respectively.
The vacuum-induced scalar field potential is defined as (up to an additive constant):
\be
\label{1-20}
	V( |\Psi |^2)	= - \frac{1}{\beta}
	\left\{ |\Psi|^2 \left[
  	 \ln{(a^3 |\Psi|^2)} - 1
  	\right]
  	+ \frac{1}{a^3}
	\right\}
	,
\ee
see Ref. \cite{Zloshchastiev:2009aw} for the detailed discussion of its properties
and related symmetry-breaking mechanism.

Further, if we choose the standard representation of the Dirac matrices
\be
	\gamma^0 =
	\begin{pmatrix}
		\textbf{1}  & 0 		\\
		0           & - \textbf{1}
	\end{pmatrix} , \
	\gamma^i =
	\begin{pmatrix}
		0           & \paul_i \\
		- \paul_i  & 0
	\end{pmatrix} ,	
\label{1-30}
\ee
where $\paul_i$ ($i=1,2,3$) are the Pauli matrices
\be
	\paul_1 =
	\begin{pmatrix}
		0 & 1 		\\
		1 & 0
	\end{pmatrix} , \
	\paul_2 =
	\begin{pmatrix}
		0 & -i \\
		i & 0
	\end{pmatrix} ,	\
	\paul_3 =
	\begin{pmatrix}
		1 & 0 \\
		0 & -1
	\end{pmatrix}
	,
\label{1-40}
\ee
then the energy density for the Lagrangian \eqref{1-10} is given by the expression
\ba
  H &=& \frac{1}{2} \left(
    {\vec E}^2 + {\vec H}^2
  \right) +
  \chi^\dagger \left[\bm{\alpha}^i \left(
    \widehat{p}_i -e A_i
  \right) + \bm{\beta} m + e \phi
  \right] \chi 
\nn\\&&	
+\frac{1}{2} \left(
    \left| \partial_t \Psi \right|^2 +
    \left| \nabla \Psi \right|^2
  \right) + V( |\Psi |^2)	,
\label{1-50}
\ea
where $\vec E, \vec H$ are electric and magnetic field's 3-vectors, 
and the matrices $\bm{\alpha}, \bm{\beta}$ are defined as
\be
	\bm\alpha^i =
	\begin{pmatrix}
		0 & \paul_i 		\\
		\paul_i & 0
	\end{pmatrix} , \
	\bm\beta =
	\begin{pmatrix}
		\textbf{1} & 0 \\
		0 & -\textbf{1}
	\end{pmatrix}
	,
\label{1-60}
\ee
where $\textbf{1}$ is a unit $2\times 2$ matrix.

The field equations can be written down as follows
\ba
&&
  \partial_\mu F^{\mu \nu} = j^\nu,
\label{2-10}\\&&
  D_\mu D^\mu \Psi + \frac{\partial \, V}{\partial \Psi^*}
  	=	\left[
  	D_\mu D^\mu + \frac{1}{\beta} \ln{(a^3 |\Psi|^2)}
  	\right] \Psi	=	0	,~~~~
\label{2-20}
\\&&
	\gamma^\mu \left(
		i \partial_\mu - e A_\mu
	\right) \chi - m \chi = 0 ,
\label{2-25}
\ea
where
\be
  j^\mu = \frac{ig}{2} \left[
		\left( D^\mu \Psi \right)^* \Psi - \Psi^* \left( D^\mu \Psi \right)
	\right] +
	e \chi^\dagger \gamma^\mu \chi
\label{2-26}
\ee
is the electric current of scalar and spinor fields.

\sscn{Angular momentum and spin observables}{sec-mo-am}

Our field configuration (represented by a regular localized solution that will be derived below),
is expected to describe a finite-size object with internal structure and field content.
Moreover, this object is supposed to
be observed by a measurement apparatus which is macroscopic and
located outside the solution's core.
Therefore,
it is necessary to define observable quantities, such as spin, in
a quantum-mechanical way.
This is going to be the main subject of this section. 

According to the interpretation of spinor wavefunctions in relativistic quantum mechanics,
an average of the physical value operator $\hat A$ 
related to the spinor part of a field solution
is given by the formula
\be\lb{e:av}
\langle A \rangle
=
\frac{\langle \chi | \hat A | \chi \rangle}{\langle \chi |  \chi \rangle}
=
\frac{\int \chi^\dagger \hat A \chi d V}
{\int \chi^\dagger \chi d V}
,
\ee
where the integration is taken over the spatial volume occupied by a system.
On the other hand,
the density of total angular momentum (TAM) of the field configuration is given by 
\be
\tamden^{\mu \nu}
=
\chi^\dagger \hat M^{\mu \nu} \chi
,
\ee
where we denoted the operator
\be
	\hat M^{\mu \nu} = x^\mu \hat p^\nu - x^\nu \hat p^\mu + 
	\frac{i}{4} \left( 
		\gamma^\mu \gamma^\nu - \gamma^\nu \gamma^\mu 
	\right) 
,
\ee
and $\hat p^\mu$ is a 4-momentum operator \cite{Berestetskii}.
In particular, we are interested in a $z$-component of TAM density,
\be\lb{e:tamden}
\tamden_z \equiv
\tamden^{1 2}
=
\chi^\dagger
\hat{M}_z
\chi
,
\ee
where 
\be
\hat{M}_z
=
\hat{M}^{1 2}
=
\hat{L}_z
+ \frac{1}{2} \hat\Sigma_3
\ee
and
\[
\hat\Sigma_3 \equiv
	\frac{i}{2} \left( 
	\gamma^1 \gamma^2 - \gamma^2 \gamma^1 
	\right) = 
\begin{pmatrix}
	\paul_3	&	0	\\
	0			&	\paul_3	
	\end{pmatrix} 
	= 
\begin{pmatrix}
	1	&	0	&	0	&	0		\\
	0	&	-1	&	0	&	0		\\
	0	&	0	&	1	&	0		\\
	0	&	0	&	0	&	-1		\\	
	\end{pmatrix} 
,
\]
and $\hat L_z = x \hat p_y - y \hat p_x = -i \partial_\varphi$,
$\varphi$ being a polar angular coordinate.
Correspondingly, the TAM's $z$-component itself is given by the formula
\be\lb{e:tam}
M_z
=
\int
\tamden_z
\,
d V 
=
\int
\chi^\dagger
\hat{M}_z
\chi
\,
d V
=
\langle \chi | \hat M_z | \chi \rangle
,
\ee
where we use the notation (\ref{e:av}) for the last step in this equation.

From the viewpoint of an external distant observer,
values $\tamden_z$ and $M_z$ are related to the spin of the field configuration
as a whole:
the observable spin (or, more precisely, a spin projection on the third axis)
can be defined as the following quantum-mechanical average
\be\lb{e:spin}
S_z
\equiv
\langle M_z \rangle
=
\frac{\langle \chi | \hat M_z | \chi \rangle}{\langle \chi |  \chi \rangle}
=
\frac{\int \chi^\dagger \hat{M}_z \chi d V}
{\int \chi^\dagger \chi d V}
= \frac{M_z} {\int \chi^\dagger \chi d V}
,
\ee
according to the definition (\ref{e:av}) and Eq. (\ref{e:tam}).\\

\sscn{Solution ansatz}{sec-mo-ss}

Using the spherical coordinates, we search for the solution in the standard form (see, for example, \cite{Berestetskii}):
\begin{eqnarray}
  \chi &=& e^{-i \epsilon t} \begin{pmatrix}
    f(r)\Omega_{jlm} \\
    (-1)^{\frac{1+l-l^\prime}{2}}h(r)\Omega_{jl^\prime m}
  \end{pmatrix},
\label{2-27}
\\
  A_{\mu} &=& \left( \phi(r),0,0,0 \right) ,
\\
  \Psi
	&=& e^{i E t}\psi(r),
\end{eqnarray}
where 
$j$ is the total angular momentum of a particle,
$l^\prime=2j-1$, $\epsilon$ and $E$ are energy-related integration constants, $\Omega_{jlm}$ is 3D spherical spinor defined as
\begin{eqnarray}
  \Omega_{l+1/2,l,m} &=&
  \begin{pmatrix}
    \sqrt{\frac{j+m}{2j}} Y_{l,m-1/2} \\
    \sqrt{\frac{j-m}{2j}} Y_{l,m+1/2}
  \end{pmatrix},
\label{2-30}
\end{eqnarray}
where
$l=j \pm 1/2$ and $m$ are the azimuthal and magnetic quantum numbers, correspondingly.
Here $Y_{l,m}$ is a spherical harmonic function normalized in the following way:
\bw
\ba
Y_{lm}(\theta,\varphi)
&=&
(-1)^{(m+|m|)/2} i^l
\left[\frac{2l+1}{4\pi}\frac{(l-|m|)!}{(l+|m|)!}\right]^{1/2}
P_{l}^{|m|}(\cos\theta)\,
e^{i m \varphi}
.
      \label{2-40}
\ea
\ew
With this ansatz in hand, 
the original field equations reduce to a set of ordinary differential equations
\begin{eqnarray}
&&
  \psi^{\prime \prime} + \frac{2}{r} \psi^\prime =
  -\left( E-g \phi \right)^2 \psi -
  \frac{1}{\beta} \psi \ln \left(a^3\psi^2 \right),
\label{2-70}\\&&
  \phi'' + \frac{2}{r}\phi^\prime =
  - 2g \left( E-g \phi \right) \psi^2 +
  e \left(
   f^2 + h^2
  \right),
\label{2-80}\\&&
  f' + \frac{1+\kappa}{r}f -
  \left( \epsilon + m - e\phi \right) h = 0,
\label{2-90}\\&&
    h' + \frac{1-\kappa}{r}h +
    \left( \epsilon - m - e \phi \right) f = 0
.
\label{2-100}
\end{eqnarray}
and $\kappa$ is a nonzero integer defined as
\be
\kappa = \begin{cases}
-(j+1/2)=-(l+1),&j=l+1/2,\\
(j+1/2)=l,&j=l-1/2,\end{cases}
\ee
where
$l$ is the orbital momentum of a charged particle.

\scn{Regular localized solution}{sec-pls}

In this section we start with rewriting the field equations in the dimensionless
form, then we proceed with the analytical derivation of the asymptotical behaviour of a regular localized solution. After that we are going to numerically obtain such a solution and study its properties.

\sscn{Dimensionless equations}{sec-dless}

If one introduces the dimensionless values
$x = r/\sqrt{\beta}$,
$\tilde{f}(x) = \beta^{1/4} a^{3/2}f(r),\
  \tilde{h}(x) = \beta^{1/4} a^{3/2}h(r)$, $\tilde{\psi} (x) =a^{3/2}\psi (r),\
  \tilde{\phi} (x) = a^{3/2} \phi (r), \
  \tilde{g}=g \sqrt{\beta / a^3}$,
$\tilde e = e \sqrt{\beta / a^3} ,
	\tilde{\epsilon} = \epsilon\sqrt{\beta},\
  \tilde{m} = m\sqrt{\beta}, \
	\tilde{E} = E\sqrt{\beta}$,
then field equations \eqref{2-70}-\eqref{2-100} take the following form:
\begin{eqnarray}
&&
    \tilde{\psi}'' + \frac{2}{x} \tilde{\psi}' =
    -\left( \tilde{E} - \tilde g\tilde{\phi} \right)^2
    \tilde{\psi} - \tilde{\psi} \ln{(\tilde\psi^2)},
\label{2-110}\\&&
    \tilde{\phi}'' + \frac{2}{x} \tilde{\phi}' = 
		- 2 \tilde{g} \left( \tilde{E} - \tilde g \tilde{\phi} \right) \tilde{\psi}^2+
    \tilde{e} \left(\tilde{f}^2 + \tilde{h}^2 \right),~~~
\label{2-120}\\&&
  \tilde{f}' + \frac{1+\kappa}{x}\tilde{f} -
	\tilde{m}_+
	\tilde{h} = 0,
\label{2-130}\\&&
  \tilde{h}' + \frac{1-\kappa}{x}\tilde{h} -
	\tilde{m}_-
	\tilde{f}
= 0
	,
\label{2-140}
\end{eqnarray}
where prime now denotes the derivation with respect to $x$, and we denoted $\tilde{m}_\pm = \tilde{m} \pm (\tilde{\epsilon} - \tilde{e} \tilde{\phi})$.
It is sometimes useful to eliminate the function $\tilde{h} (x)$ from the equations above, then the equation \eqref{2-140} can be replaced with
\bw
\ba
&&
\tilde{f}''
+
\left(
\frac{2}{x}
+
\frac{\tilde{e} \tilde{\phi}'}{\tilde{m}_+}
\right)
\tilde{f}'
-
\left(
\tilde{m}_+ \tilde{m}_-
-
\frac{1+\kappa}{x}
\frac{\tilde{e} \tilde{\phi}'}{\tilde{m}_+}+\frac{\kappa(1+\kappa)}{x^2}
\right)
\tilde{f}
= 0,
~~~~~
\lb{e:eomf}
\ea
\ew
whereas an expression for $\tilde{h} (x)$ can be found in an algebraic
way from \eqref{2-130}, as soon as the solutions for $\tilde{f} (x)$ and $\tilde{\phi} (x)$ have been obtained.

Furthermore, in the dimensionless notations the energy density \eqref{1-50}
acquires the following form:
\ba
\tilde{H}
&=&
	\frac{1}{2} \left(
    \tilde{\phi}^{\prime 2}
		+
		\tilde{\psi}^{\prime 2}
		+
		\tilde E^2 \tilde \psi^2
  \right)+
    \tilde h \tilde f^\prime -
    \tilde f \tilde h^\prime - \frac{2}{r} \tilde f \tilde h 
\nn\\&&		
+
    \tilde m \left( \tilde f^2 - \tilde h^2 \right)
    + \tilde e \tilde \phi \left( \tilde f^2 + \tilde h^2 \right)
\nn\\&&
	- {\tilde \psi}^2 \left(
    \ln {\tilde \psi}^2 - 1
  \right)
,
\label{2-190}
\ea
whereas the dimensionless charge density \eqref{2-26} becomes
\begin{equation}
  \tilde \rho \equiv {\tilde j}^0 = \tilde g \tilde E {\tilde \psi}^2 +
  \tilde e \left( {\tilde f}^2 +{\tilde h}^2 \right)
.
\label{2-210}
\end{equation}

Once these densities are computed, analytically or numerically,
the dimensionless values of the total rest mass-energy and total charge
can be computed, respectively, as
\be
\restm
=
 4 \pi \int \limits_0^\infty
    \tilde H x^2 dx
,
\ \
\tilde Q =
 4 \pi \int \limits_0^\infty
  \tilde \rho\, x^2  dx
.
\ee
Apart from these observables, we would also be interested
in the behavior of observable values related to the solution's angular momentum.
Substituting Eq. (\ref{2-27}) for a ground state,
given by the expression
\be
  \chi_{(0)} = e^{-i \epsilon_0 t} 
	\begin{pmatrix}
    f(r)\\
		0\\
    h(r)\cos\theta\\
		h(r)\sin\theta \,e^{i \varphi}
  \end{pmatrix}
	,
\ee
into Eq. (\ref{e:tamden}),
we obtain the dimensionless density of the TAM's $z$-component
\be\lb{e:dlesstamden}
\tilde\tamden_z = 
\frac{1}{2}
\left(\tilde f^2 + \tilde h^2\right)
,
\ee
whereas the dimensionless form of the
TAM's  $z$-component itself is
\be\lb{e:dlesstam}
\tilde M_z = 
2 \pi
\int\limits_0^\infty
\left(
\tilde f^2 + \tilde h^2
\right)
x^2 d x
,
\ee
according to Eq. (\ref{e:tam}).

Notice that,
since (a dimensionless form of) the norm in our case can be computed as 
\be\lb{e:dlessnrm}
\langle \tilde\chi |  \tilde\chi \rangle = 4 \pi \int\limits_0^\infty
\left(\tilde f^2 + \tilde h^2\right) x^2 d x
= 2 \tilde M_z
,
\ee
we can immediately derive 
the value of the quantum-mechanical spin of our solution.
Using Eqs. (\ref{e:spin}), (\ref{e:dlesstam}) and (\ref{e:dlessnrm}),
we obtain
\be
S_z
=
\frac{\tilde M_z}{\langle \tilde\chi |  \tilde\chi \rangle}
= 
\frac{1}{2}
,
\ee
as expected.

\sscn{Asymptotics}{sec-asymp}

Using the field equations \eqref{2-110}-\eqref{e:eomf} one can show that
the components of scalar field $\tilde{\psi}$, scalar EM potential $\tilde{\phi}$
and spinor fields $\tilde{f}$ and $\tilde{h}$, behave at large $x$ in the following way:
\ba
  \tilde{\psi}(x) &\propto&
	\psi_\infty e^{-\frac{x^2}{2}},
 \\
 \tilde{\phi}(x) &\propto& \tilde{\phi}_\infty -\frac{\tilde Q}{x}
,
\label{2-146}\\
  \tilde{f}(x) &\propto&
  f_\infty \frac{e^{-\mu x}}{x},
\\
  \tilde{h}(x) &\propto&
  - f_\infty \sqrt{\frac{\tilde{m} - \tilde{\e} + \tilde{e} \tilde \phi_\infty}{\tilde{m} + \tilde{\e} - \tilde{e} \tilde \phi_\infty}}
  \frac{e^{- \mu x}}{x} ,
\ea
where
$\mu = \sqrt{\tilde m^2-\left( \tilde\e - \tilde e \tilde \phi_\infty \right)^2}$, and $f_\infty, \psi_\infty$ and
$\tilde{\phi}_\infty = \tilde{\phi} (\infty)$ are the constants that can be fixed by boundary conditions.

At $x \to 0$, functions 
can be expanded into the Taylor series
\begin{eqnarray}
  \tilde{f}(x) &=& \tilde{f}_0 + \tilde{f}_1 x + \ldots ,\
  \tilde{h}(x)=\tilde{h}_1 x + \tilde{h}_3\frac{x^3}{3!} + \ldots ,\;
\nn\\
  \tilde{\psi}(x) &=& \tilde{\psi}_0 + \tilde{\psi}_2\frac{x^2}{2} + \ldots ,
\
  \tilde{\phi}(x) = \tilde{\phi}_0 + \tilde{\phi}_2 \frac{x^2}{2} + \ldots ,
\nn
\end{eqnarray}
which ensures that their  behavior is nonsingular in the origin as well.

These asymptotic expansions analytically prove the existence of regular localized solution at small and large
values of $x$ on a positive semi-axis.
This gives us a strong evidence for an existence of the solution that is not only regular for non-negative values of $x$ but also localized, i.e., its field components all vanish at spatial infinity.

\sscn{Limit cases}{sec-limc}

Apart from asymptotic properties of solutions of Eqs. (\ref{2-120})-(\ref{e:eomf}),
it is instructive to study a few limit cases when some of the charge parameters of the theory
are either small or large.

\textit{Small} $\tilde e$.
In this case, the spinor part of the field equations begins to decouple from the electrostatic and scalar parts.
Indeed, in the limit  $\tilde e \to 0$ 
we can neglect the last term in Eq. \eqref{2-120}, then Eqs. (\ref{2-110})-(\ref{2-140}) become, respectively,
\begin{eqnarray}
&&
    \tilde{\psi}'' + \frac{2}{x} \tilde{\psi}' =
    -\left( \tilde{E} - \tilde g\tilde{\phi} \right)^2
    \tilde{\psi} - \tilde{\psi} \ln{(\tilde\psi^2)},
\label{2-110se}\\&&
    \tilde{\phi}'' + \frac{2}{x} \tilde{\phi}' = - 2 \tilde{g} \left( \tilde{E} - \tilde g \tilde{\phi} \right) \tilde{\psi}^2,
\label{2-120se}\\&&
  \tilde{f}' + \frac{1+\kappa}{x}\tilde{f} -
	(\tilde{m} + \tilde{\epsilon})
	\tilde{h} = 0,
\label{2-130se}\\&&
  \tilde{h}' + \frac{1-\kappa}{x}\tilde{h} -
	(\tilde{m}- \tilde{\epsilon})
	\tilde{f} = 0,
\label{2-140se}
\end{eqnarray}
so one can see that the former two equations evolve independently from the latter two.
In other words, we are arriving at the two independent systems: the former 
has been studied in past \cite{Dzhunushaliev:2012zb}, 
and the latter 
is equivalent to a system of free spinor fields,
which is known for not possessing regular localized solutions.
Therefore, it is crucial to have $\tilde e \not= 0$, which maintains the coupling of spinors
to the scalar EM potential and hence to the scalar $\psi$.

\textit{Large} $\tilde g$.
In this case, in order to maintain a structure of our eigenvalue problem ($\tilde E$ should not
drop off our equations when making approximations),
we perform the rescaling $\tilde{\phi} \to \tilde{\Phi} = \tilde g \tilde{\phi}$.
Then Eqs. (\ref{2-110})-(\ref{2-140}) become
\ba
&&
    \tilde{\psi}'' + \frac{2}{x} \tilde{\psi}' =
    -\left( \tilde{E} - \tilde{\Phi} \right)^2
    \tilde{\psi} - \tilde{\psi} \ln{(\tilde\psi^2)}
		,
\\&&
    \tilde{\Phi}'' + \frac{2}{x} \tilde{\Phi}' = 
		- 2 \tilde{g}^2 \left( \tilde{E} -  \tilde{\Phi} \right) \tilde{\psi}^2+
    \tilde{e} \tilde{g} \left(\tilde{f}^2 + \tilde{h}^2 \right)
		,~~~
\\&&
  \tilde{f}' + \frac{1+\kappa}{x}\tilde{f} -
\left(\tilde{m} + \tilde{\epsilon} -  \frac{\tilde{e}}{\tilde{g}} \tilde{\Phi}  \right)
	\tilde{h} = 0
	,
\\&&
  \tilde{h}' + \frac{1-\kappa}{x}\tilde{h} 
- \left(\tilde{m} -\tilde{\epsilon} +     \frac{\tilde{e}}{\tilde{g}} \tilde{\Phi}  \right)
	\tilde{f}
= 0
.
\ea
One can see that in the limit $|\tilde g | \to \infty$ 
the latter two equations become equal to Eqs. (\ref{2-130se})-(\ref{2-140se}),
which are known for not having solutions which are both finite in the origin and vanishing at 
spatial infinity.
Therefore, when increasing a magnitude of $\tilde g $ one should expect the shrinking
of a set of parameters and boundary conditions which are allowed for existence of regular localized
solutions.
In the next section, we will find some numerical confirmation of this statement.

\begin{figure}
\centering
\subfloat[]{
  \includegraphics[width=\sct\columnwidth]{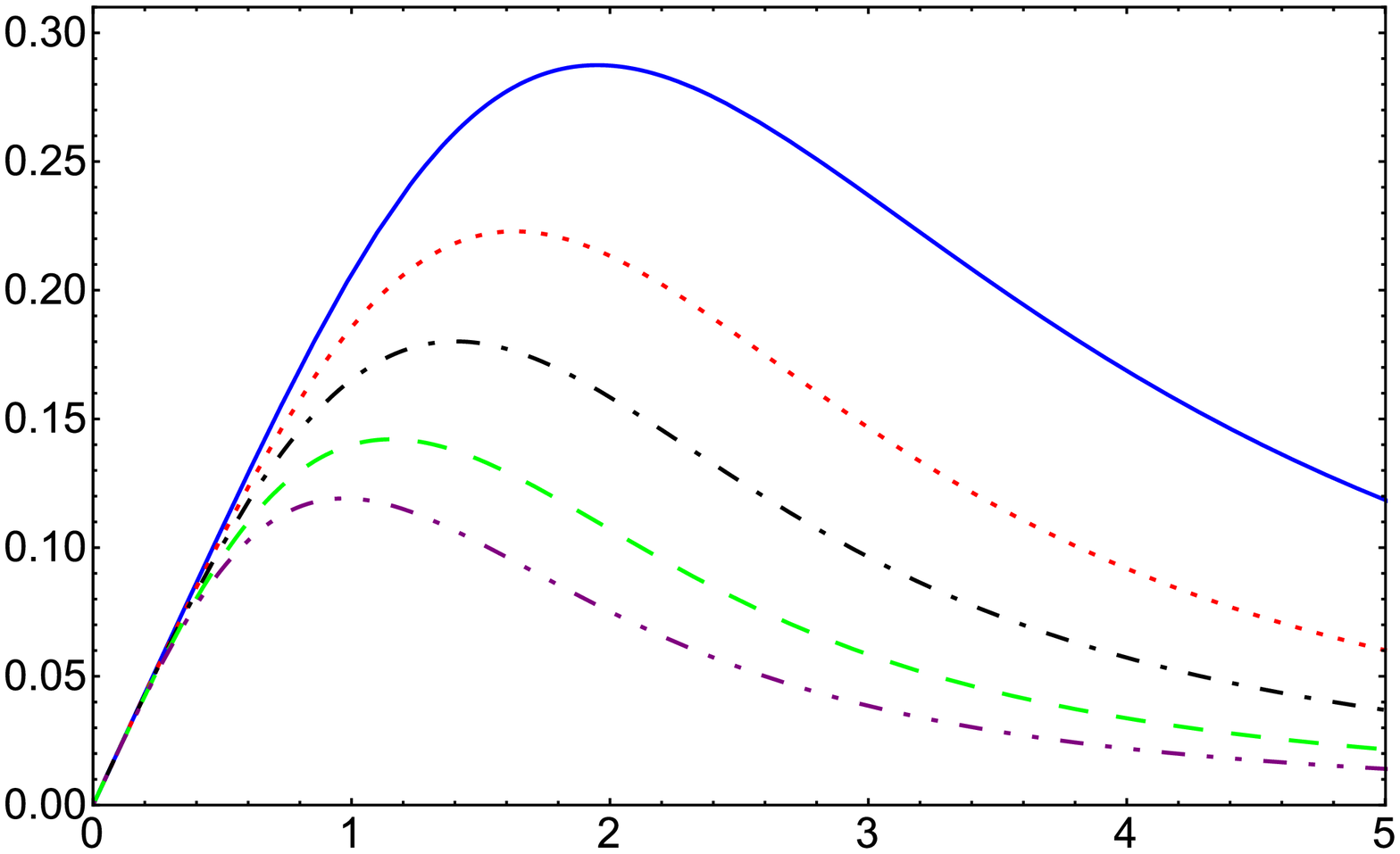}
}
\subfloat[]{
  \includegraphics[width=\sct\columnwidth]{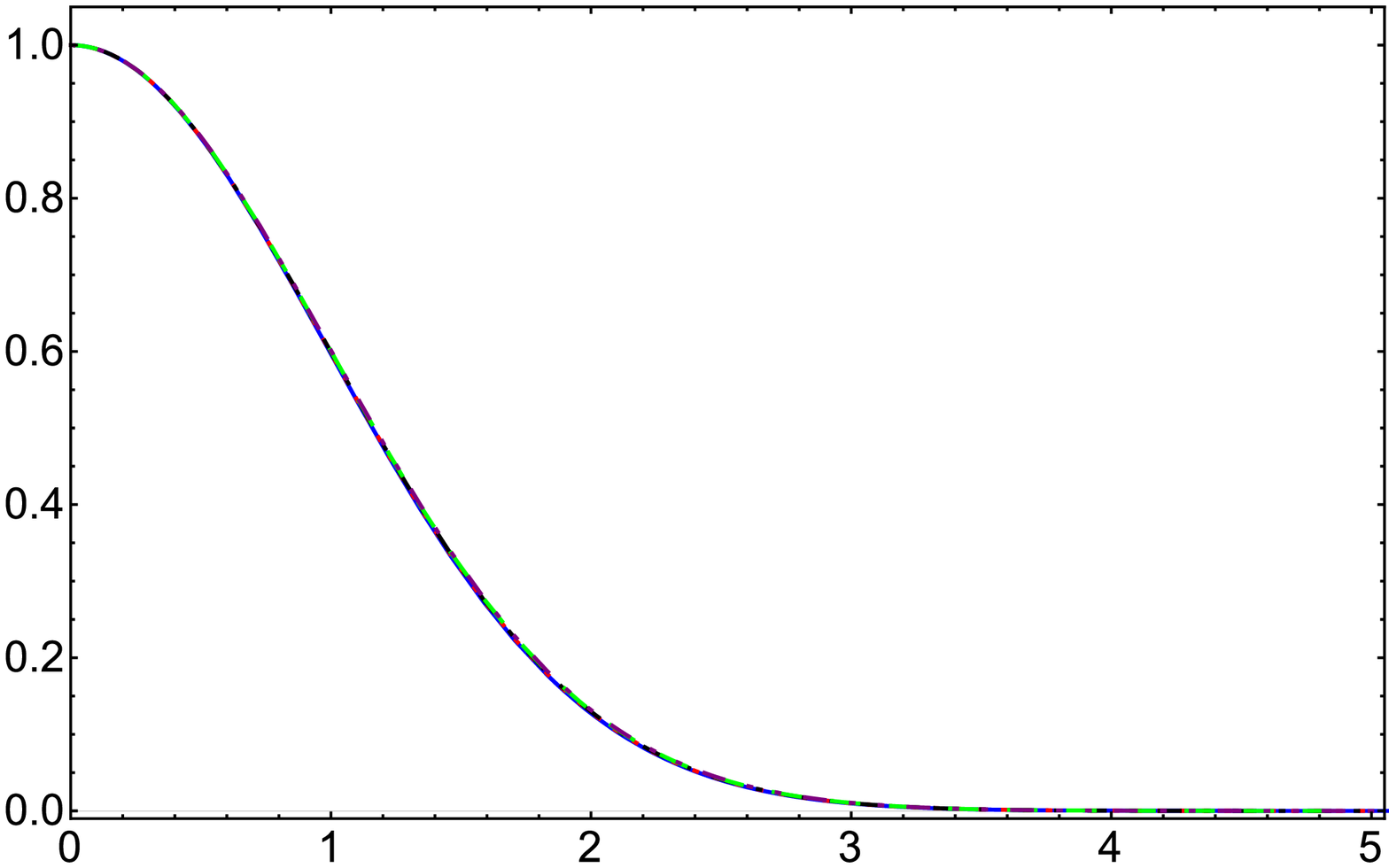}
}
\hspace{0mm}
\subfloat[]{
  \includegraphics[width=\sct\columnwidth]{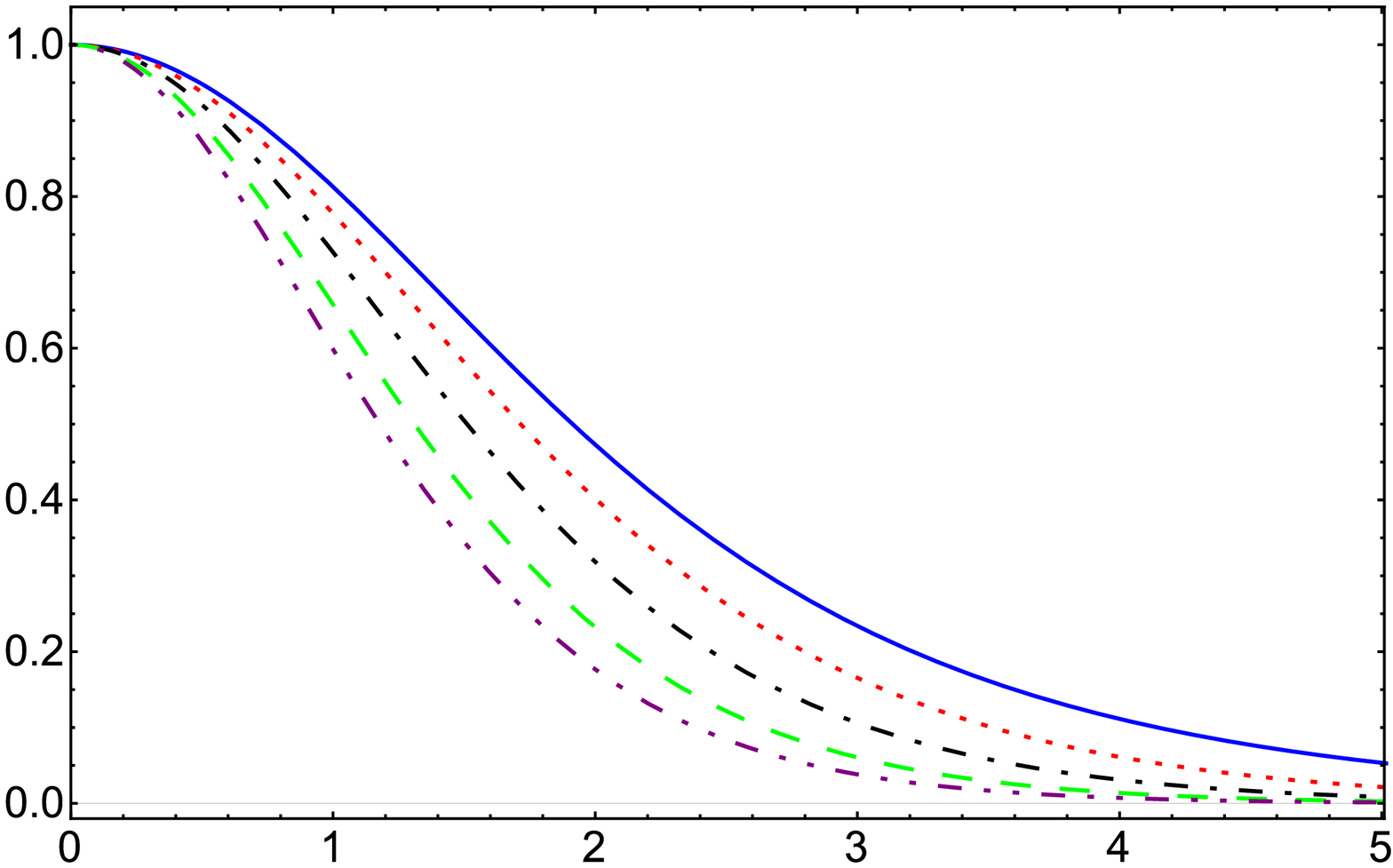}
}
\subfloat[]{
  \includegraphics[width=\sct\columnwidth]{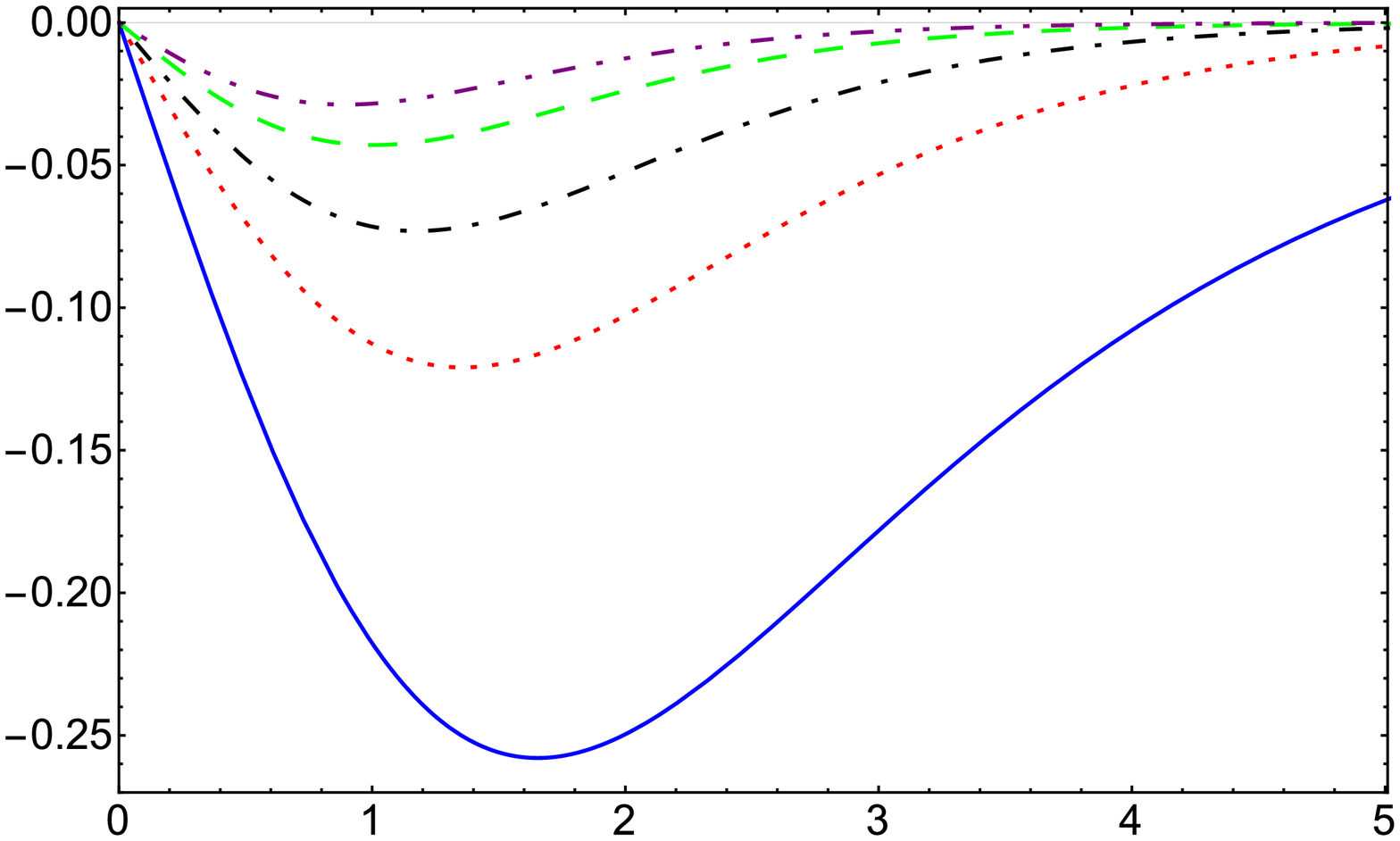}
}
\caption{Electrostatic field $\tilde{\phi}'$ (\textbf{a}), scalar field $\tilde{\psi}$ (\textbf{b}), 
spinor field components $\tilde{f}$ (\textbf{c}) and $\tilde{h}$ (\textbf{d}), versus the dimensionless distance from the origin, evaluated at 
$\tilde{g}=0.1$, $\tilde \phi_0=1$, $\tilde e=1,\;\tilde{\psi}_0= 1,\;\tilde{f}_0=1$,
and the following values of $\tilde{m}$: $0.4$ (solid curve), $1.5$ (dotted curve),
$3.0$ (dash-dotted curve),
$6.0$ (dashed curve) and $10.0$ (dash-double dotted curve). 
}
\label{f-fields1}
\end{figure}

\sscn{Numerical solution}{sec-sol}

To begin with, note that the dimensionless field equations contain terms with $x$ in denominators. 
Therefore, we have to carry out our computations starting from some small number $x = \d$ (here we choose $\d=0.001$).
Next, using solutions of Eqs.  \eqref{2-110}-\eqref{2-140}
in the form of the Taylor series expanded near the origin, we obtain the following
boundary conditions:
\ba
&&
    \tilde{\psi}(\d) =\tilde{\psi}_0 +
    \tilde{\psi_2}\frac{\delta^2}{2},\;
    \tilde{\psi}'(\d)=\tilde{\psi_2} \delta,\nn\\&&
    \tilde{\phi}(\d)=\tilde{\phi}_0 +
    \tilde{\phi_2}\frac{\delta^2}{2}
		,
    \tilde{\phi}'(\d)= \tilde{\phi_2} \delta ,
\label{2-160}\\&&
    \tilde{f}(\d) = \tilde{f}_0,\;
    \tilde{h}(\d)=\tilde h_1 \delta
.
\nn
\ea
As long as for our calculations we assume spherical symmetry, we can impose that
\be\lb{e:kappas}
\kappa=-1
.
\ee
A numerical solution of Eqs. \eqref{2-110}-\eqref{2-140} is obtained by solving an eigenvalue problem with the parameters $\tilde E$ and $\tilde\e$ chosen to be eigenvalues. 
We can represent a set of these eigenvalues of system of equations \eqref{2-110}-\eqref{2-140} as a point in the parametric space which coordinates are eigenvalue parameters themselves, namely: $\tilde{\e}$, $\tilde{E}$. 
We presume that this parametric space contains continuous subsets of such points, therefore once we find any set of eigenvalues (as a point of such a subset), we can then retrieve other solutions. 
Therefore, we must somehow find this subset of the parametric space for we initially set the parameters of our problem to some trial values (\textit{e.g.}, $\tilde{\e}=1$ and $\tilde{E}=1$).
For instance, one can assume:
\begin{eqnarray}
\tilde{m}=1,\;\tilde \phi_0=1,\;\tilde e=1,\;\tilde{\psi}_0= 1,\;\tilde{f}_0=1,
\label{2-170}
\end{eqnarray}
whereas for the parameter $\tilde{g}$
we consider the following three cases:

\begin{figure}
\centering
\subfloat
{
  \includegraphics[width=\sc\columnwidth]{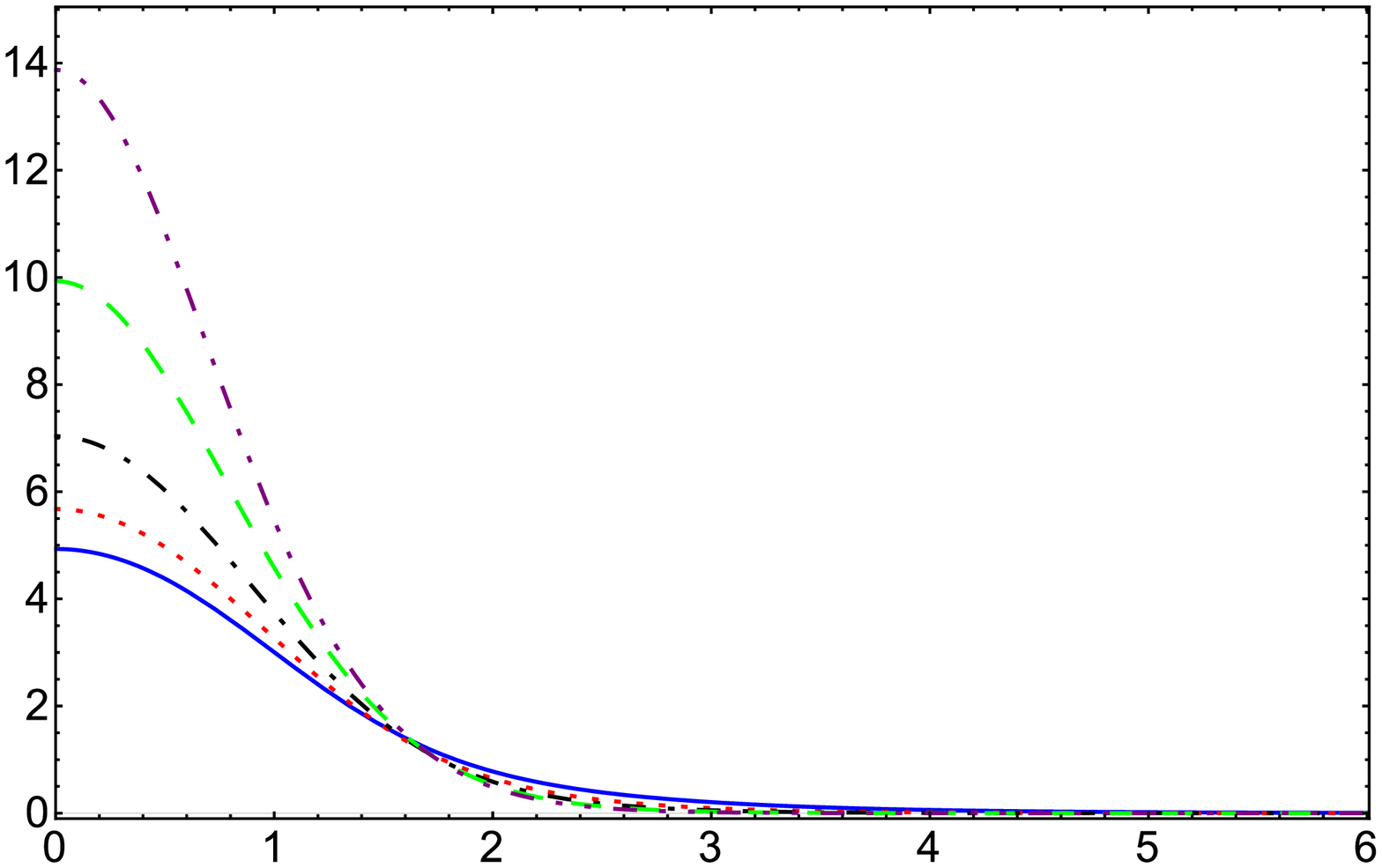}
}
\hspace{0mm}
\subfloat
{
  \includegraphics[width=\sc\columnwidth]{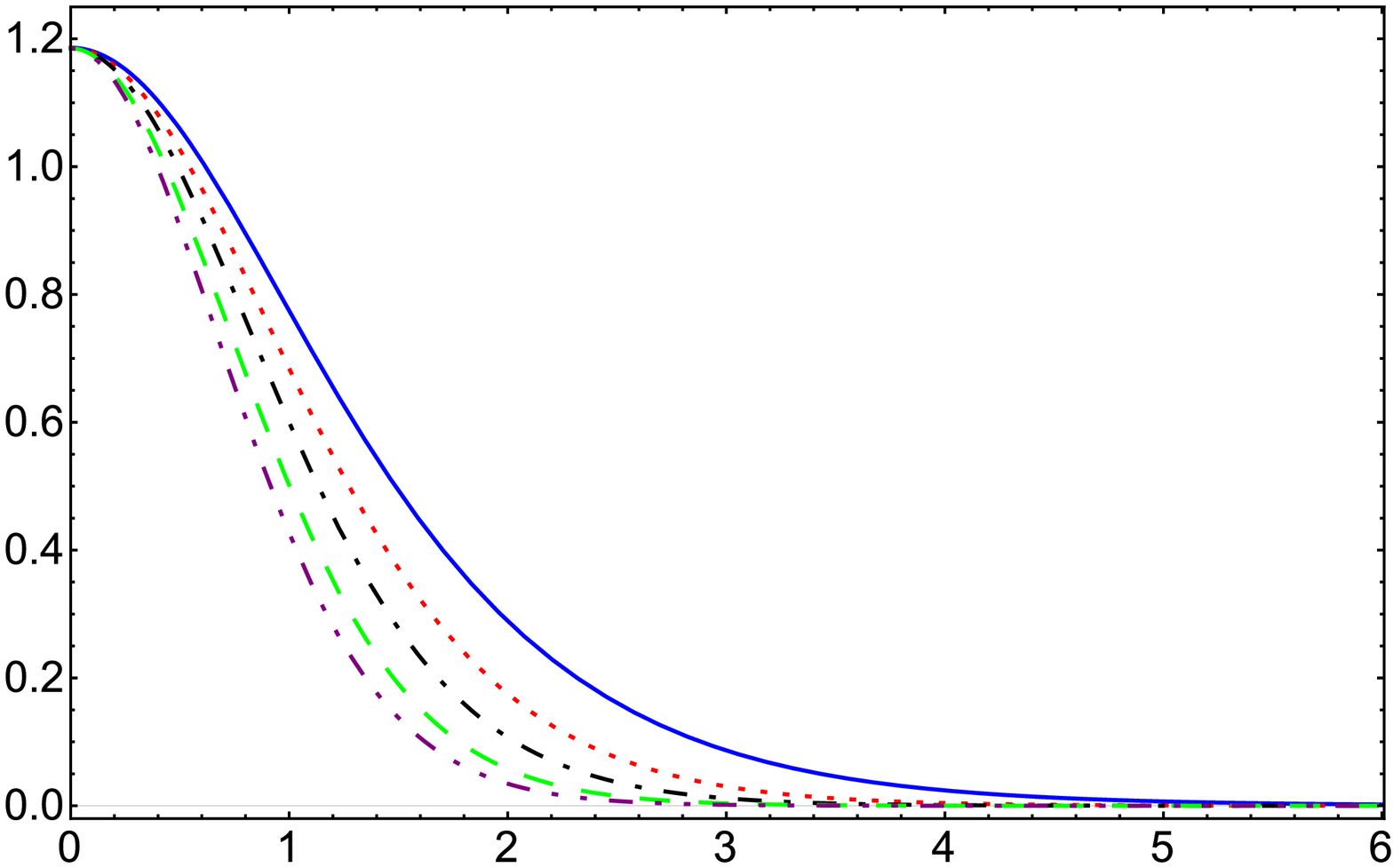}
}
\hspace{0mm}
\subfloat
{
  \includegraphics[width=\sc\columnwidth]{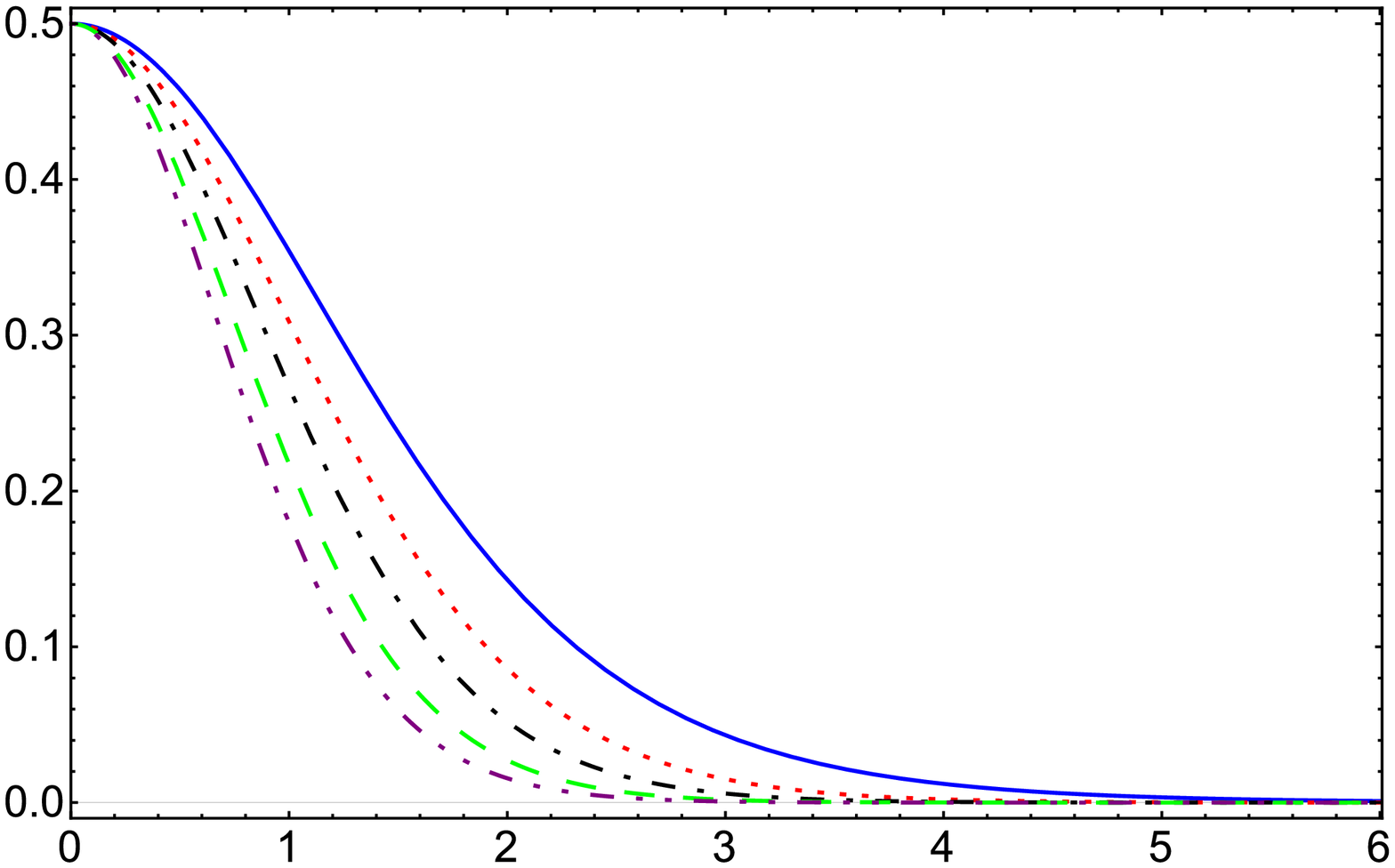}
}
\caption{Dimensionless energy density $\tilde H$ (upper panel), charge density $\tilde\rho$  (middle panel) and TAM's $z$-component density $\tilde\tamden_z$ (lower panel), 
versus the dimensionless 
distance from the origin, evaluated at 
$\tilde{g}=0.1$, $\tilde \phi_0=1$, $\tilde e=1,\;\tilde{\psi}_0= 1,\;\tilde{f}_0=1$,
and the following values of $\tilde{m}$: $0.4$ (solid curve), $1.5$ (dotted curve),
$3.0$ (dash-dotted curve),
$6.0$ (dashed curve) and $10.0$ (dash-double dotted curve). 
}
\label{f-ench1}
\end{figure}

\begin{figure}[htbt]
\begin{center}\epsfig{figure=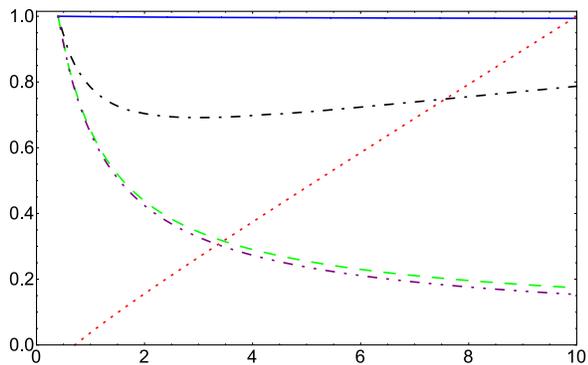,width=1.0\columnwidth}\end{center}
\caption{
Cumulative plot of an eigenvalue $\tilde{E}$ (solid curve, vertical axis' scale 1:1.86),
eigenvalue $\tilde{\epsilon}$ (dotted curve,
scale 1:9.81),
total rest mass-energy $\restm$ (dot-dashed curve,
scale 1:121.5), 
total charge $\tilde{Q}$ (dashed curve,
scale 1:42.1),
and
TAM's $z$-component $ \tilde{M}_z $ (double-dot-dashed curve,
scale 1:20.55),
versus $\tilde{m}$.
Plotted values correspond to solutions of Eqs. \eqref{2-110}-\eqref{2-140},
evaluated
at $\tilde{g}=0.1,\;\tilde \phi_0=1,\;\tilde e=1$, $\tilde{\psi}_0= 1,\;\tilde{f}_0=1$.
}
\label{f-cum1}
\end{figure}

\subsubsection{Case $\tilde{g} \ll 1$}

Let us assume
$\tilde{g}=0.1$ for definiteness.
We start numerical computations by
shooting in different directions from a point 
$\{ 	\tilde{\e}=1,\;\tilde{E}=1 \} $
in the parametric space.
Here by shooting we mean trying to solve equations with intuitively guessed sets of numbers.
We can simplify a task of finding a point of the eigenvalue subset by dividing it into the following two tasks.
The first one is searching for eigenvalues of the system \eqref{2-110}-\eqref{2-120}, by using shooting method in one-dimensional space of eigenvalue parameter $\tilde{E}$. 
The second task is searching for eigenvalues of the system \eqref{2-130}-\eqref{2-140} by means of the same shooting method in one-dimensional space of the parameter $\tilde{\e}$. 
These two systems of equations still depend on each other, therefore, in order to begin calculations, we assume the trial values for the first set of functions, then we calculate the second one.

Thus, we begin by assuming a simplest possible ansatz, 
$
\tilde{f}(x)=0,\;\tilde{h}(x)=0,
$
and calculate $\tilde \psi(x)$ and $\tilde \phi(x)$ at the first step, while setting $\tilde{E}$ to appropriate value.
Then, at the second step, we solve the system \eqref{2-130}-\eqref{2-140} by changing value of $\tilde{\e}$ and using the function $\tilde \phi(x)$  which has been calculated in the previous step. 
After second step of calculations is completed, 
we return to the first step with non-zero functions $\tilde f(x)$ and $\tilde h(x)$; this process repeats until we get good convergence of the eigenvalues.

Proceeding that way, we eventually obtain
\begin{eqnarray}
\tilde{\e} \approx 2.55,\;\tilde{E} \approx 1.85815
.
\end{eqnarray}
It is worth noting that, while Eq. \eqref{2-146} implies that $\tilde{\phi}(x) \to 0$ at spatial infinity, the numerical solution of Eqs. \eqref{2-110}-\eqref{2-120} asymptotically tends to a constant $\tilde \phi_\infty$. 
A nonzero value of $\phi_\infty$ is a consequence of gauge invariance of field equations: the function $\phi(x)$  is defined up to an additive constant $\phi_\infty$.

Thus, in this section the system \eqref{2-110}-\eqref{2-140} has been numerically solved as a non-linear eigenvalue problem for the parameters $\tilde E$ and $\tilde \e$.
The corresponding plots of fields for several regular localized solutions are given in Figs. \ref{f-fields1}, 
whereas the dimensionless energy density \eqref{2-190}, charge density \eqref{2-210} 
and TAM density \eqref{e:dlesstamden} 
are plotted in Fig. \ref{f-ench1}.
Computations also show that $\mu$ 
in this case takes real values only for $\tilde{m}\gtrsim 0.4$.

Furthermore, in Fig. \ref{f-cum1} we present a cumulative plot of different values, which characterize our solution, versus dimensionless bare spinor mass $\tilde m$. 
One can see that the eigenvalue $\tilde E$ varies very slowly, the total rest mass-energy has a local minimum near $\tilde m \approx 3$ and grows with increasing bare spinor mass, whereas the dimensionless charge decreases with increasing bare spinor mass. 
The latter means that the ratio $\tilde Q/\restm$ decreases
with increasing bare spinor mass, which is usually the case 
for composite objects with a non-compensated half-integer spin.

\begin{figure}
\centering
\subfloat[]{
  \includegraphics[width=\sct\columnwidth]{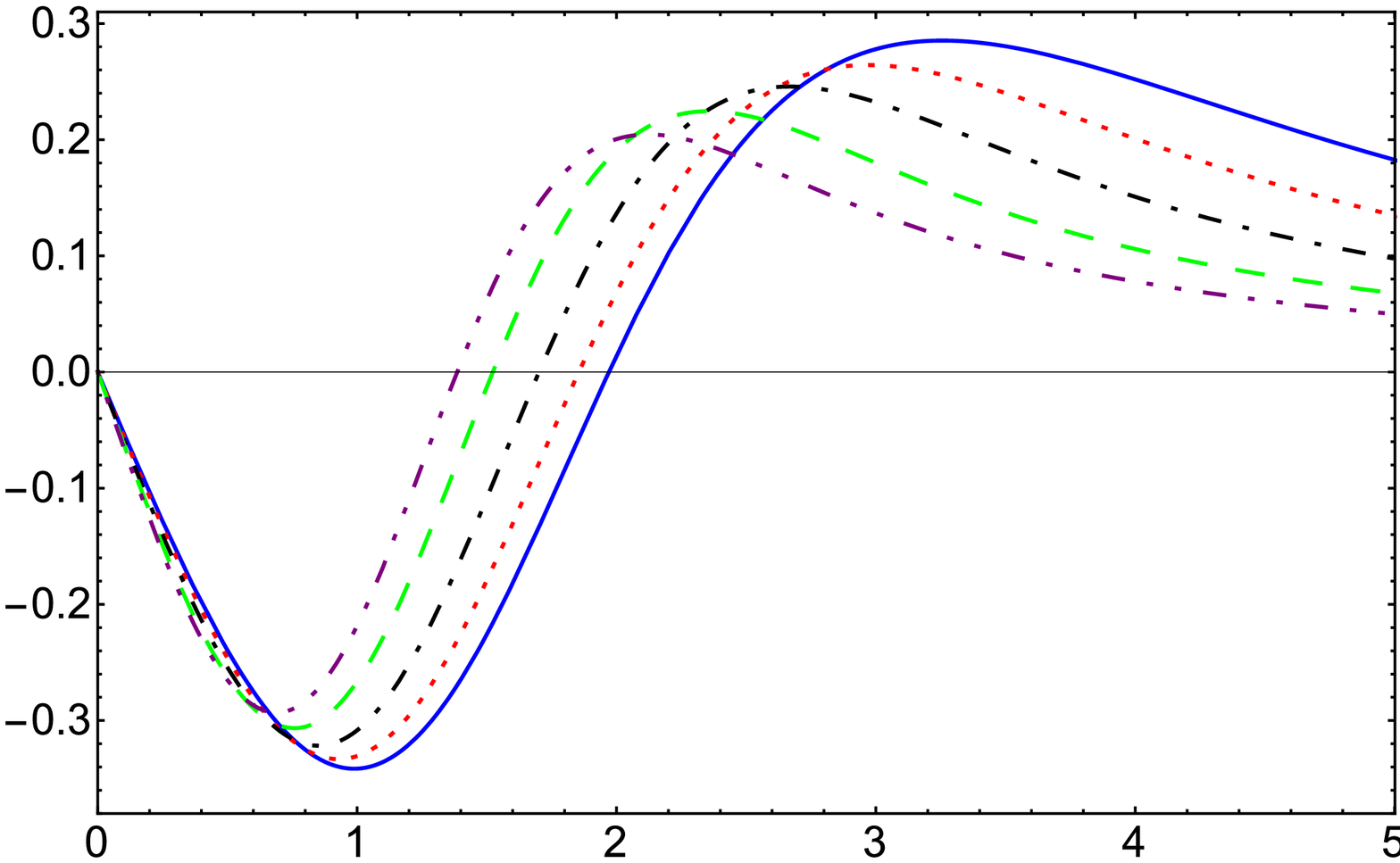}
}
\subfloat[]{
  \includegraphics[width=\sct\columnwidth]{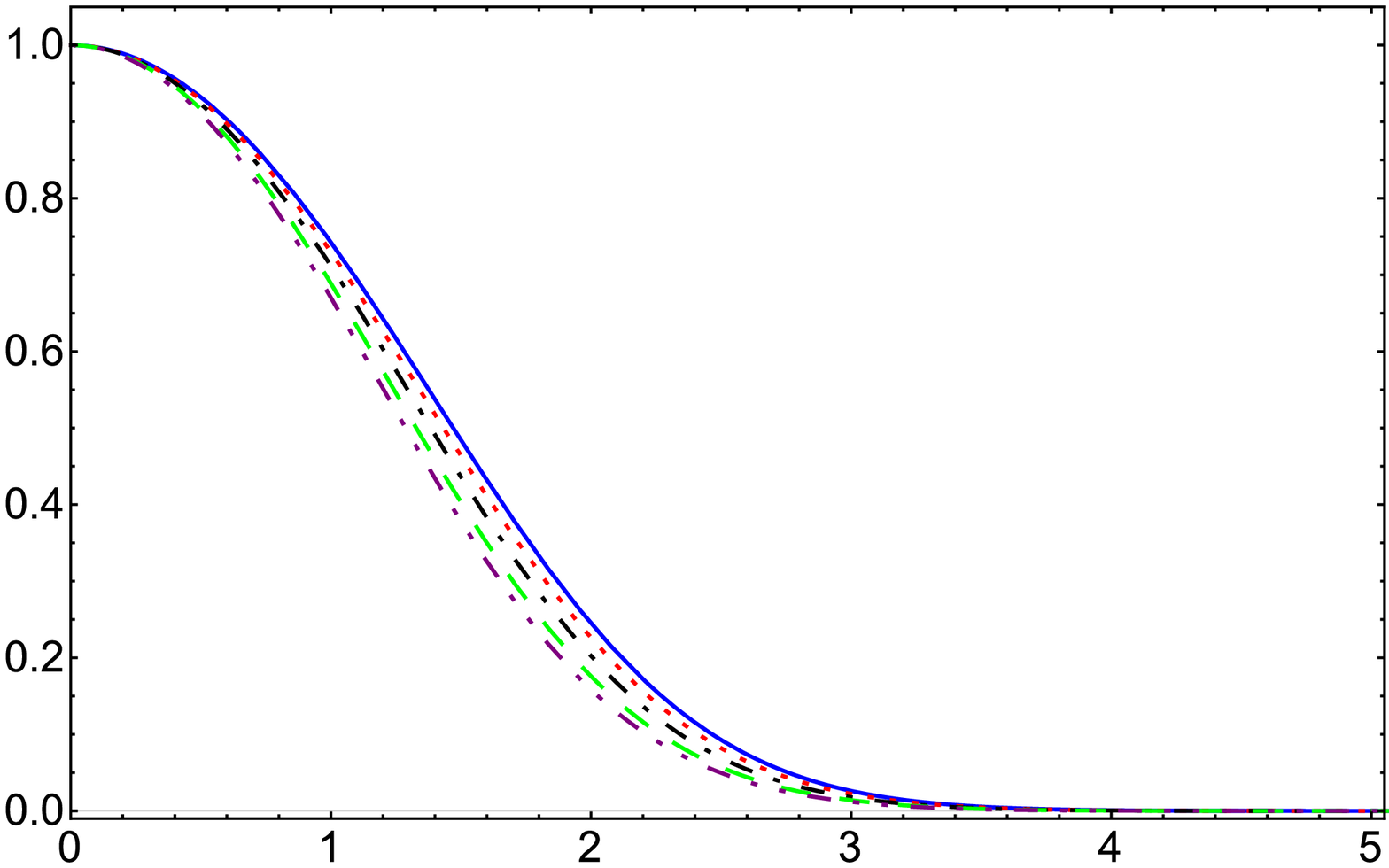}
}
\hspace{0mm}
\subfloat[]{
  \includegraphics[width=\sct\columnwidth]{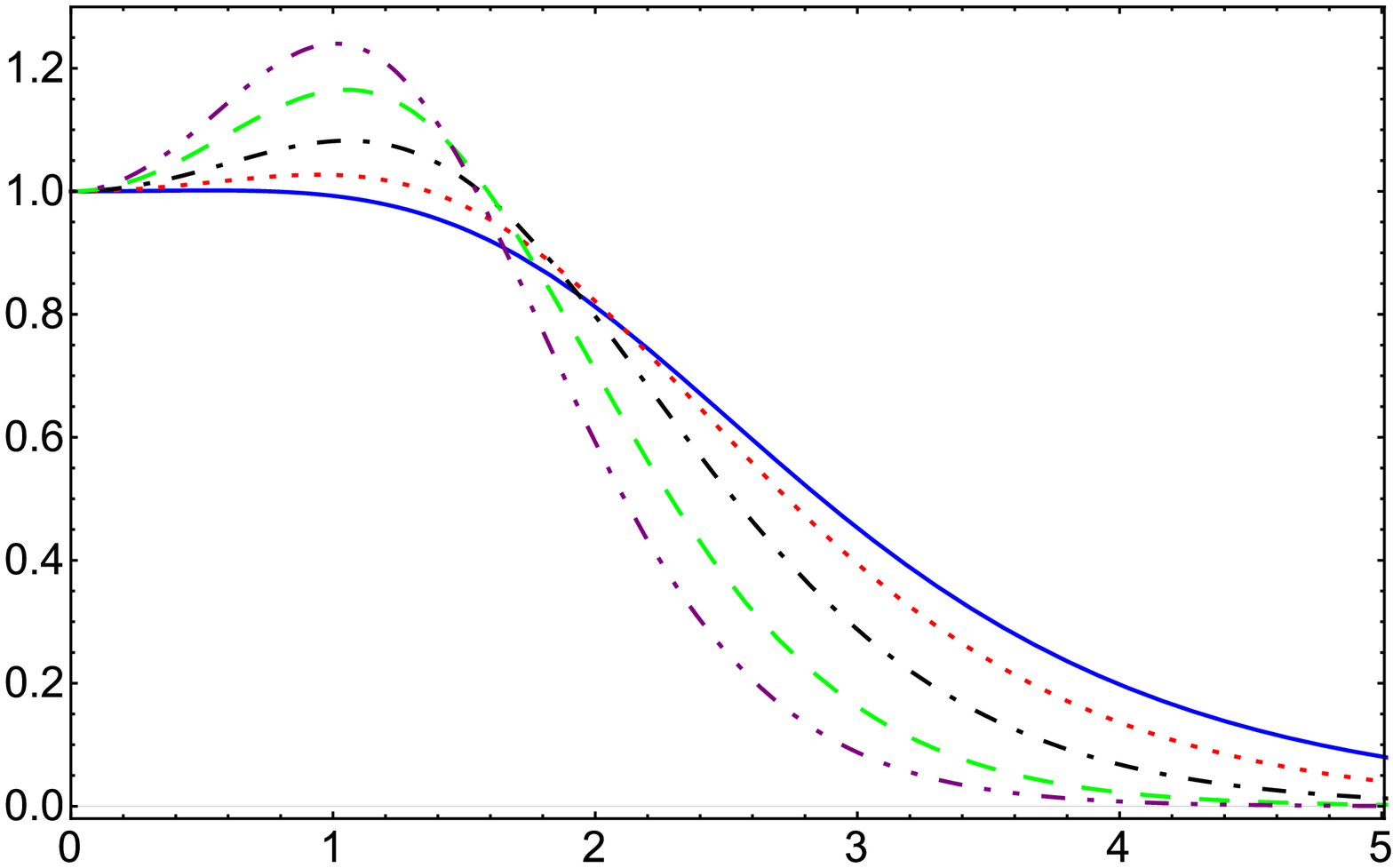}
}
\subfloat[]{
  \includegraphics[width=\sct\columnwidth]{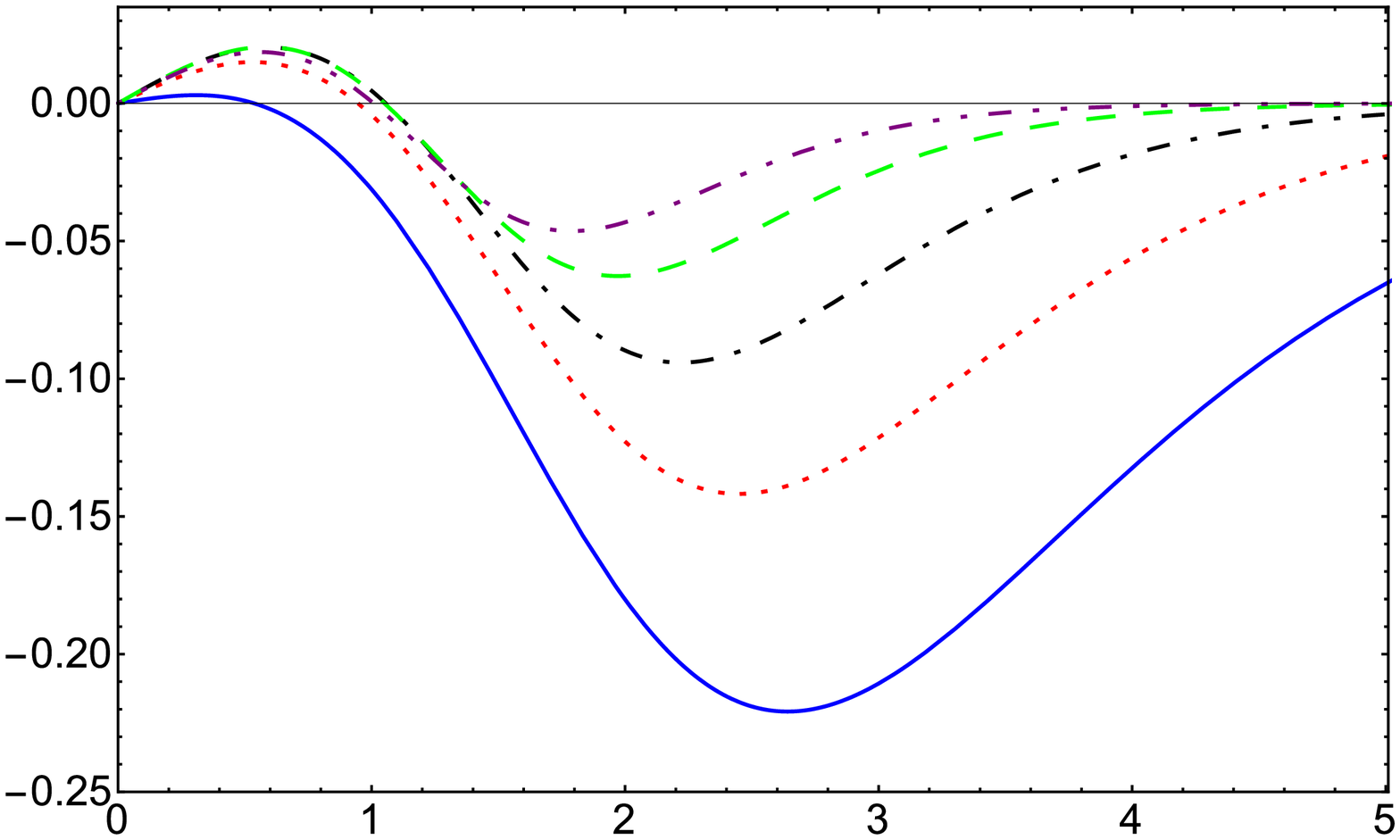}
}
\caption{Electrostatic field $\tilde{\phi}'$ (\textbf{a}), scalar field $\tilde{\psi}$ (\textbf{b}), 
spinor field components $\tilde{f}$ (\textbf{c}) and $\tilde{h}$ (\textbf{d}), versus the dimensionless distance from the origin, evaluated at 
$\tilde{g}=1$, $\tilde \phi_0=1$, $\tilde e=1,\;\tilde{\psi}_0= 1,\;\tilde{f}_0=1$,
and the following values of $\tilde{m}$: $0.7$ (solid curve), $1.5$ (dotted curve),
$3.0$ (dash-dotted curve),
$6.0$ (dashed curve) and $10.0$ (dash-double dotted curve).
}
\label{f-fields2}
\end{figure}

\subsubsection{Case $\tilde{g} \sim 1$}

Let us set now
$\tilde{g}=1$ for definiteness.
Similarly to the previous case, we set corresponding parameters to the following values
\begin{eqnarray}
\tilde \phi_0=1,\;\tilde e=1,\;\tilde{\psi}_0= 1,\;\tilde{f}_0=1.
\label{2-180}
\end{eqnarray}
As in the previous case, we find that there exists a similar constraint upon $\mu$: 
it takes real values only for $\tilde{m}\gtrsim 0.7$.

\begin{figure}
\centering
\subfloat
{
  \includegraphics[width=\sc\columnwidth]{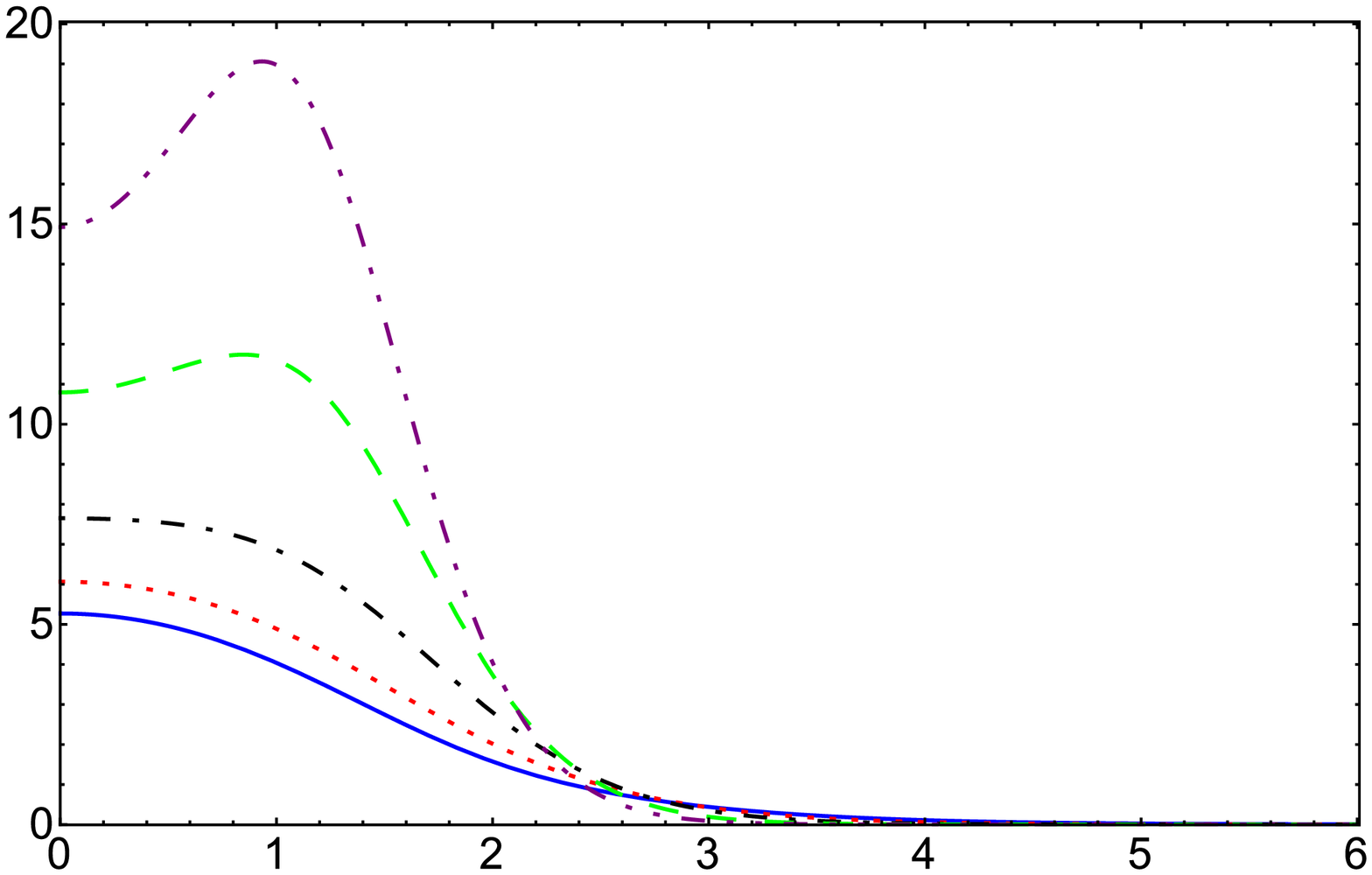}
}
\hspace{0mm}
\subfloat
{
  \includegraphics[width=\sc\columnwidth]{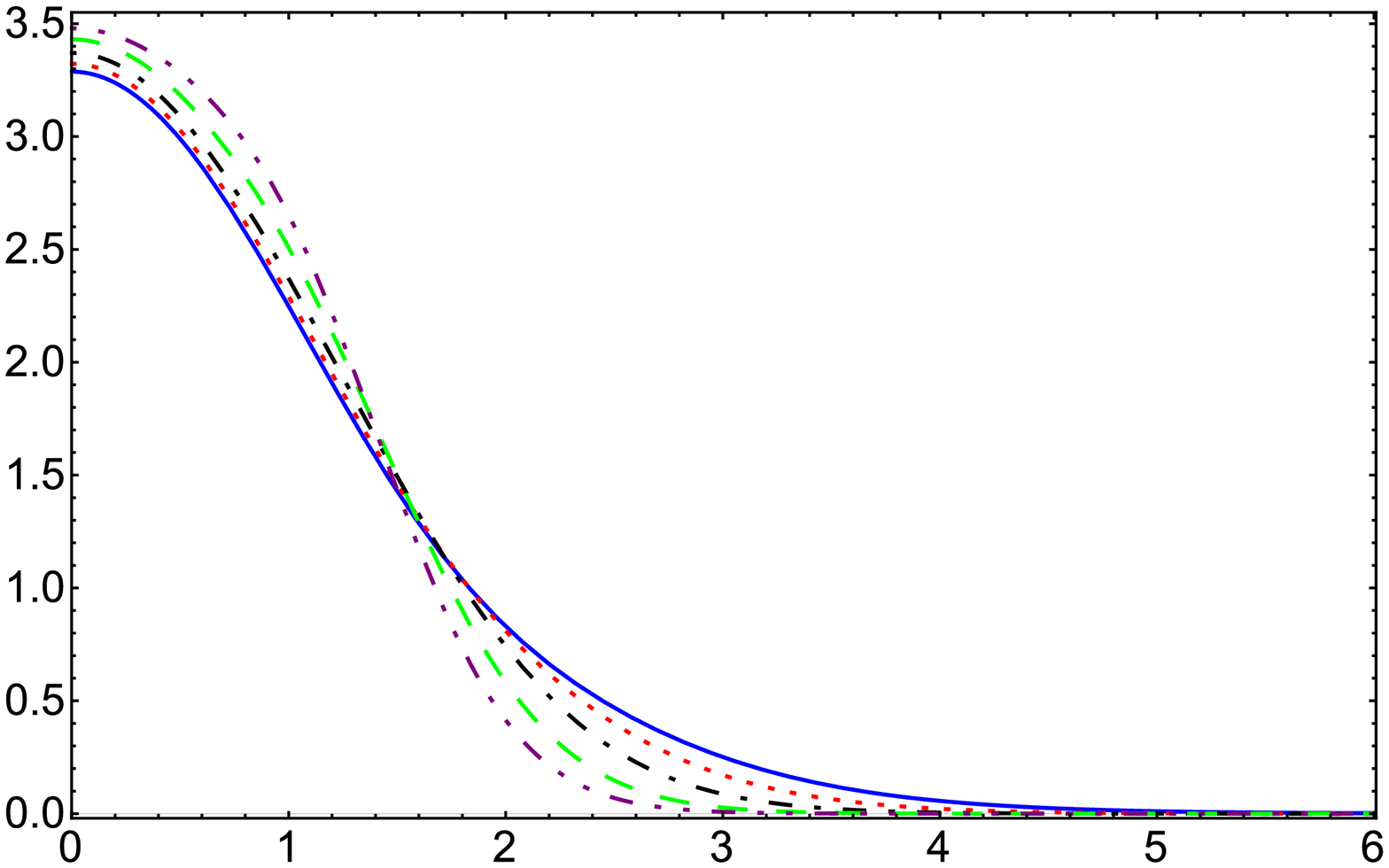}
}
\hspace{0mm}
\subfloat
{
  \includegraphics[width=\sc\columnwidth]{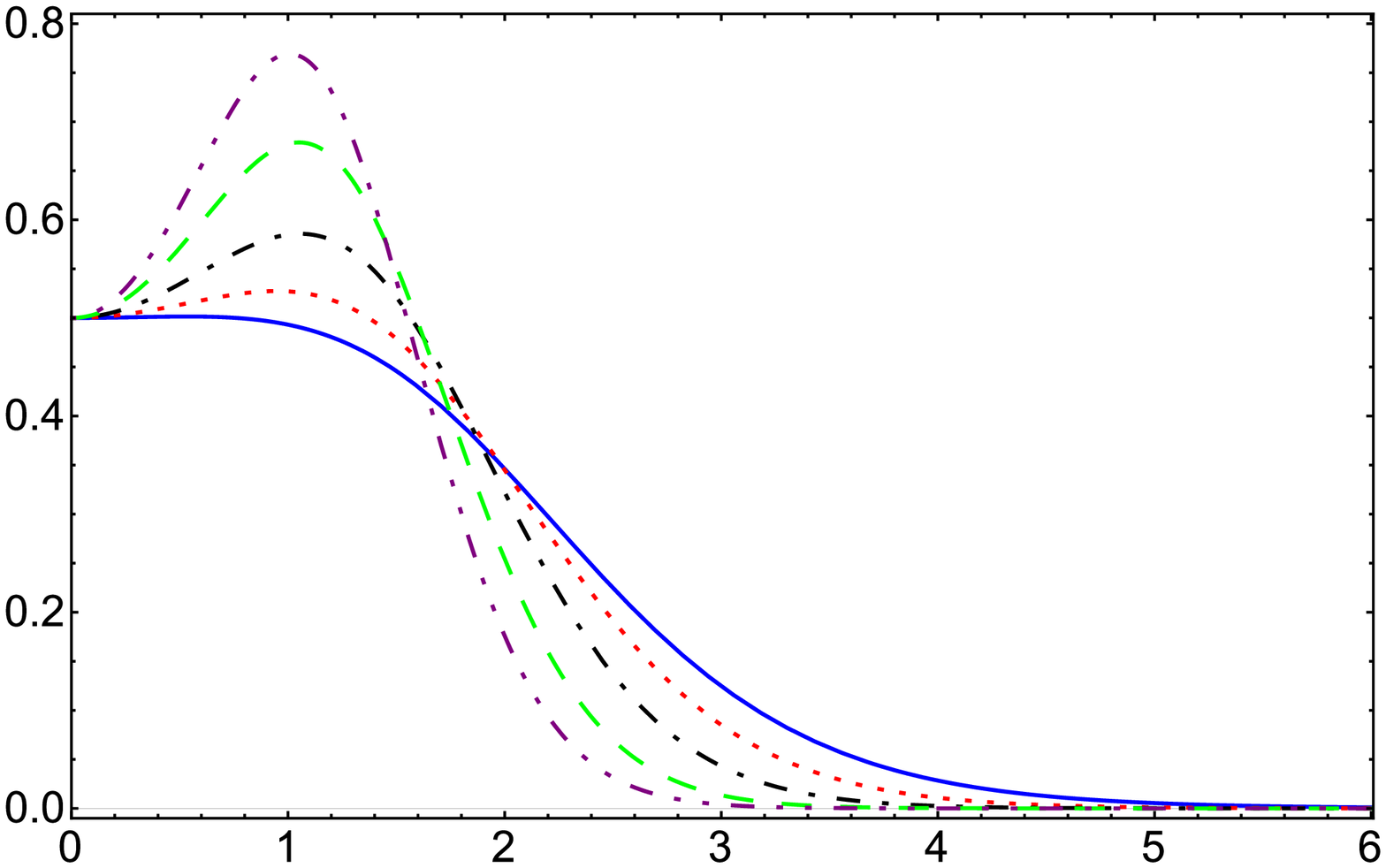}
}
\caption{Dimensionless energy density $\tilde H$ (upper panel), charge density $\tilde\rho$  (middle panel) and TAM's $z$-component density $\tilde\tamden_z$ (lower panel),
versus the dimensionless 
distance from the origin, evaluated at 
$\tilde{g}=1$, $\tilde \phi_0=1$, $\tilde e=1,\;\tilde{\psi}_0= 1,\;\tilde{f}_0=1$,
and the following values of $\tilde{m}$: $0.7$ (solid curve), $1.5$ (dotted curve),
$3.0$ (dash-dotted curve),
$6.0$ (dashed curve) and $10.0$ (dash-double dotted curve). 
}
\label{f-ench2}
\end{figure}

\begin{figure}[htbt]
\begin{center}\epsfig{figure=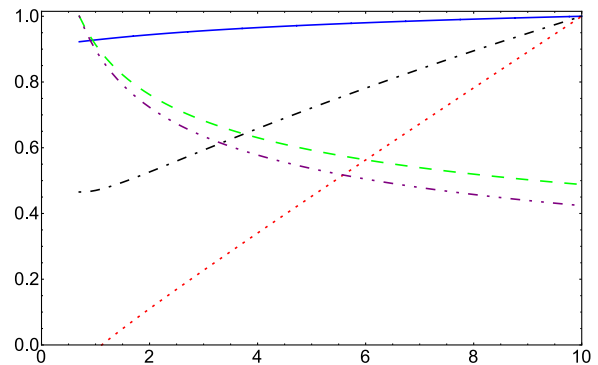,width=1.0\columnwidth}\end{center}
\caption{
Cumulative plot of the eigenvalue $\tilde{E}$ (solid curve, vertical axis' scale 1:2.48),
eigenvalue $\tilde{\epsilon}$ (dotted curve, scale 1:9.43),
total rest mass-energy $\restm$ (dot-dashed curve, scale 1:461.1), 
total charge $\tilde{Q}$ (dashed curve,
scale 1:42.1),
and
TAM's $z$-component $\tilde{M}_z$ (double-dot-dashed curve,
scale 1:45.6),
versus $\tilde{m}$.
Plotted values correspond to solutions of Eqs. \eqref{2-110}-\eqref{2-140}, evaluated
at $\tilde{g}=1$, $\tilde \phi_0=1$, $\tilde e=1,\;\tilde{\psi}_0= 1,\;\tilde{f}_0=1$. }
\label{f-cum2}
\end{figure}

Field profiles of the  regular localized solutions found are shown in Fig. \ref{f-fields2}. 
The dimensionless energy density \eqref{2-190}, charge density \eqref{2-210} 
and TAM density \eqref{e:dlesstamden} are shown in Fig. \ref{f-ench2}. 
Finally, the cumulative plot of characteristic values is given in Fig. \ref{f-cum2}.
It is qualitatively similar 
to the previous case (cf. Fig. \ref{f-cum1}), 
except that the local minimum of function $\restm (\tilde m)$ has shifted
towards smaller values of the bare spinor mass $\tilde m$.
This can be explained by when $\tilde{g}$ is increasing then a larger proportion of mass-energy
comes from a non-spinor part, therefore the value $\restm$ as a function of $\tilde m$
grows faster.

\subsubsection{Case $\tilde{g} > 1$}\lb{sec-sol-lg}


If one adopts the following set of calculation parameters
\begin{eqnarray}
\tilde \phi_0=1,\;\tilde e=1,\;\tilde{\psi}_0= 1,\;\tilde{f}_0=1, \;\tilde{m}=8.7
,
\label{2-190a}
\end{eqnarray}
one can find regular localized solutions at $ 1 < \tilde{g} \leqslant 4$, whereas 
for $\tilde{g} > 4$ no such solutions have thus far been found.
Moreover, no regular localized solutions were found for some other trial sets of calculation parameters. While no conclusive statements can be made at this stage, this probably indicates that a set of allowed calculation parameters for a regular localized solution is shrinking with increasing magnitude of $\tilde{g}$. 
This confirms the analysis done in Sec. \ref{sec-limc} above.

\scn{Conclusion}{sec-con}

In this work, we proposed a model of an electrically charged fermion as a self-consistent solution of electromagnetic and spinor fields interacting with the physical vacuum effectively described by the logarithmic superfluid.

In the case of spherical symmetry, the asymptotic behavior of the electrical potential for our solution reveals that at distances which are large in comparison with the size of the solution's core, the electrical field is indistinguishable from the Coulomb one. This was confirmed both analytically and numerically. The exponential (Gaussian) decay of the spinor and scalar field components of our solution imply that these fields are detectable only in the vicinity of the core, but become technically non-observable (at least, in the leading-order approximation) at distances which are large compared to the size of the core.
This essentially means that these fields can be regarded as the internal degrees of freedom which can be probed only at a sufficiently large scale of energy and momentum. 

Furthermore, the self-energy and total charge of this object have also been numerically computed. These turn out to be finite, similar to those of the case  \cite{Dzhunushaliev:2012zb}.
Therefore, the rest mass of a fermion arises as a result of interaction of gauge fields 
with the physical vacuum. 

Thus, a distant observer would see the object described by our solution as a half-integer spin particle with mass and charge, whereas its internal degrees of freedom can be probed only at sufficiently large scales of energy and momentum. Therefore, apart from the conventional Fermi particles, our solution can find applications in theory of exotic localized objects, such as gauged fermionic $Q$-balls which would be spinor analogues of the gauged $Q$-balls
discussed in Refs. 
\cite{gul14,gul15}. 

Finally, we hypothesize, based on results of Secs. \ref{sec-limc} and \ref{sec-sol-lg},
that regular localized solutions of this system cease to exist at very large $\tilde g$'s or very small $\tilde e$'s.
This can  explain, at least qualitatively,
why for known interactions the $U (1)$ gauge couplings never reach values which are either too large or too small.

\begin{acknowledgments}
V.D. acknowledges Grant No. $\Phi.0755$  in fundamental research in Natural Sciences by the Ministry of Education and Science of the Republic of Kazakhstan.
The research of K.Z., as well as a visit of V.D. to Durban University of Technology,
were supported by the National Research Foundation of South Africa under Grant No. 98892.
Proofreading of the manuscript by P. Stannard is greatly appreciated.
\end{acknowledgments}

\def\AnP{Ann. Phys.}
\def\APP{Acta Phys. Polon.}
\def\CJP{Czech. J. Phys.}
\def\CMPh{Commun. Math. Phys.}
\def\CQG {Class. Quantum Grav.}
\def\EPL  {Europhys. Lett.}
\def\IJMP  {Int. J. Mod. Phys.}
\def\JMP{J. Math. Phys.}
\def\JPh{J. Phys.}
\def\FP{Fortschr. Phys.}
\def\GRG {Gen. Relativ. Gravit.}
\def\GC {Gravit. Cosmol.}
\def\LMPh {Lett. Math. Phys.}
\def\MPL  {Mod. Phys. Lett.}
\def\Nat {Nature}
\def\NCim {Nuovo Cimento}
\def\NPh  {Nucl. Phys.}
\def\PhE  {Phys.Essays}
\def\PhL  {Phys. Lett.}
\def\PhR  {Phys. Rev.}
\def\PhRL {Phys. Rev. Lett.}
\def\PhRp {Phys. Rept.}
\def\RMP  {Rev. Mod. Phys.}
\def\TMF {Teor. Mat. Fiz.}
\def\prp {report}
\def\Prp {Report}

\def\jn#1#2#3#4#5{{#1}{#2} {\bf #3}, {#4} {(#5)}} 

\def\boo#1#2#3#4#5{{\it #1} ({#2}, {#3}, {#4}){#5}}




\end{document}